\documentclass[manuscript,times]{aastex62}
\usepackage{booktabs}

\shorttitle{}
\shortauthors{Jiang et al.}

\usepackage{subfigure}
\usepackage{url} 
\usepackage{amsmath}
\usepackage{multirow}
\usepackage{graphicx} 
\usepackage{epstopdf}
\begin{document}

%
%


\title{Whistler Waves As a Signature of Converging Magnetic Holes in Space Plasmas}

%
%

\correspondingauthor{Wence Jiang, Hui Li}
\email{joe.jiang@ucl.ac.uk}

\author[0000-0001-7431-5759]{Wence Jiang}
\affiliation{State Key Laboratory of Space Weather, National Space Science Center, CAS, Beijing, China}
\affiliation{University of Chinese Academy of Sciences, Beijing, China}
\affiliation{Mullard Space Science Laboratory, University College London, Dorking RH5 6NT, UK}

\author[0000-0002-0497-1096]{Daniel Verscharen}
\affiliation{Mullard Space Science Laboratory, University College London, Dorking RH5 6NT, UK}
\email{d.verscharen@ucl.ac.uk}

\author[0000-0002-4839-4614]{Hui Li}
\affiliation{State Key Laboratory of Space Weather, National Space Science Center, CAS, Beijing, China}
\affiliation{University of Chinese Academy of Sciences, Beijing, China}
\email{hli@nssc.ac.cn}

\author[0000-0001-6991-9398]{Chi Wang}
\affiliation{State Key Laboratory of Space Weather, National Space Science Center, CAS, Beijing, China}
\affiliation{University of Chinese Academy of Sciences, Beijing, China}
\email{cw@spaceweather.ac.cn}

\author[0000-0001-6038-1923]{Kristopher G. Klein}
\affiliation{Department of Planetary Sciences, University of Arizona, Tucson, AZ, USA}
\email{kgklein@arizona.edu}


\begin{abstract}

Magnetic holes are plasma structures that trap a large number of particles in a magnetic field that is weaker than the field in its surroundings. The unprecedented high time-resolution observations by NASA's Magnetospheric Multi-Scale (MMS) mission enable us to study the particle dynamics in magnetic holes in the Earth's magnetosheath in great detail. {We reveal the local generation mechanism of whistler waves by a combination of Landau-resonant and cyclotron-resonant wave-particle interactions of electrons in response to the large-scale evolution of a magnetic hole.} As the magnetic hole converges, a pair of counter-streaming electron beams form near the hole's center as a consequence of the combined action of betatron and Fermi effects. The beams trigger the generation of slightly-oblique whistler waves. Our conceptual prediction is supported by a remarkable agreement between our observations and numerical predictions from the Arbitrary Linear Plasma Solver (ALPS). Our study shows that wave-particle interactions are fundamental to the evolution of magnetic holes in space and astrophysical plasmas.

\end{abstract}

\section{Introduction}\label{sec0}

Space and astrophysical plasmas exhibit electromagnetic fluctuations and inhomogeneous structures across a very broad range of scales  \citep{Schekochihin2009,Alexandrova2013,Verscharen2019b}. In the Earth's magnetosheath, plasma turbulence and coherent structures are abundant as a consequence of the plasma's bow-shock crossing and processes around the magnetopause such as magnetic reconnection and field draping  \citep{Retino2007,Tsurutani2011,Karimabadi2014}. 

Magnetic holes are an important type of spatially non-uniform and non-linear coherent plasma structure. They are characterised by a local dip in the magnetic field with an anti-correlation between density and magnetic field variations. They occur in the terrestrial magnetosheath  \citep{Tsurutani1982,Fazakerley1994,Cattaneo1998,Sahraoui2006,Yao2020} and other space plasmas like the solar wind  \citep{Tsurutani2011}, the heliosheath  \citep{Burlaga2006} and cometary environments  \citep{Russell1987,Plaschke2018}. A possible mechanism for the creation of magnetic holes is the mirror-mode instability, which is a non-propagating electromagnetic plasma instability  \citep{Chandrasekhar1958,Hasegawa1969,Kaufmann1970,Tsurutani1982,Southwood1993,Fazakerley1994,Fazakerley1995,Kuznetsov2007,Soucek2008,Kunz2014}. Alternative generation mechanisms include solitons \citep{Baumgartel1999,Li2016}, phase-steepened Alfv{\'e}n waves \citep{Tsurutani2002} and decaying turbulence \citep{Haynes2015}.

The spatial scale of magnetic holes in the Earth's magnetosheath varies from 10$\rho_\mathrm{e}$ ($\approx$ 10 km) to 5-40$\rho_\mathrm{p}$ ($\approx$ 500-3000 km)  \citep{Tsurutani2011,Yao2020,Liu2020}, where $\rho_\mathrm{e}$ and $\rho_\mathrm{p}$ are the gyroradii of the electrons and protons. Magnetic holes are capable of trapping particles due to the mirror force from their non-uniform magnetic field. These trapped particles bounce back and forth between the mirror points of these structures. At small scales, kinetic effects such as micro-instabilities can regulate the dynamics of both the trapped and the untrapped particles in magnetic holes via wave-particle interactions. If a magnetic hole changes in depth, betatron and type-1 Fermi acceleration cause the particles to evolve collectively in velocity space  \citep{Southwood1993,Pantellini1995,Kivelson1996,Chisham1998,Soucek2011,Ahmadi2018,Breuillard2018,Li2021}. However, there is still a significant lack of direct evidence for particle diffusion in wave-particle interactions at kinetic scales and of the understanding of the multi-scale evolution of magnetic holes. Here, we present such direct evidences for these processes and the important role of wave-particle interactions in a converging magnetic hole based on MMS multi-spacecraft data.

\section{Particle diffusion in converging magnetic holes: a multi-scale model}\label{sec1}

\begin{figure*}[!htbp]
\centering
\includegraphics[width=0.9\textwidth]{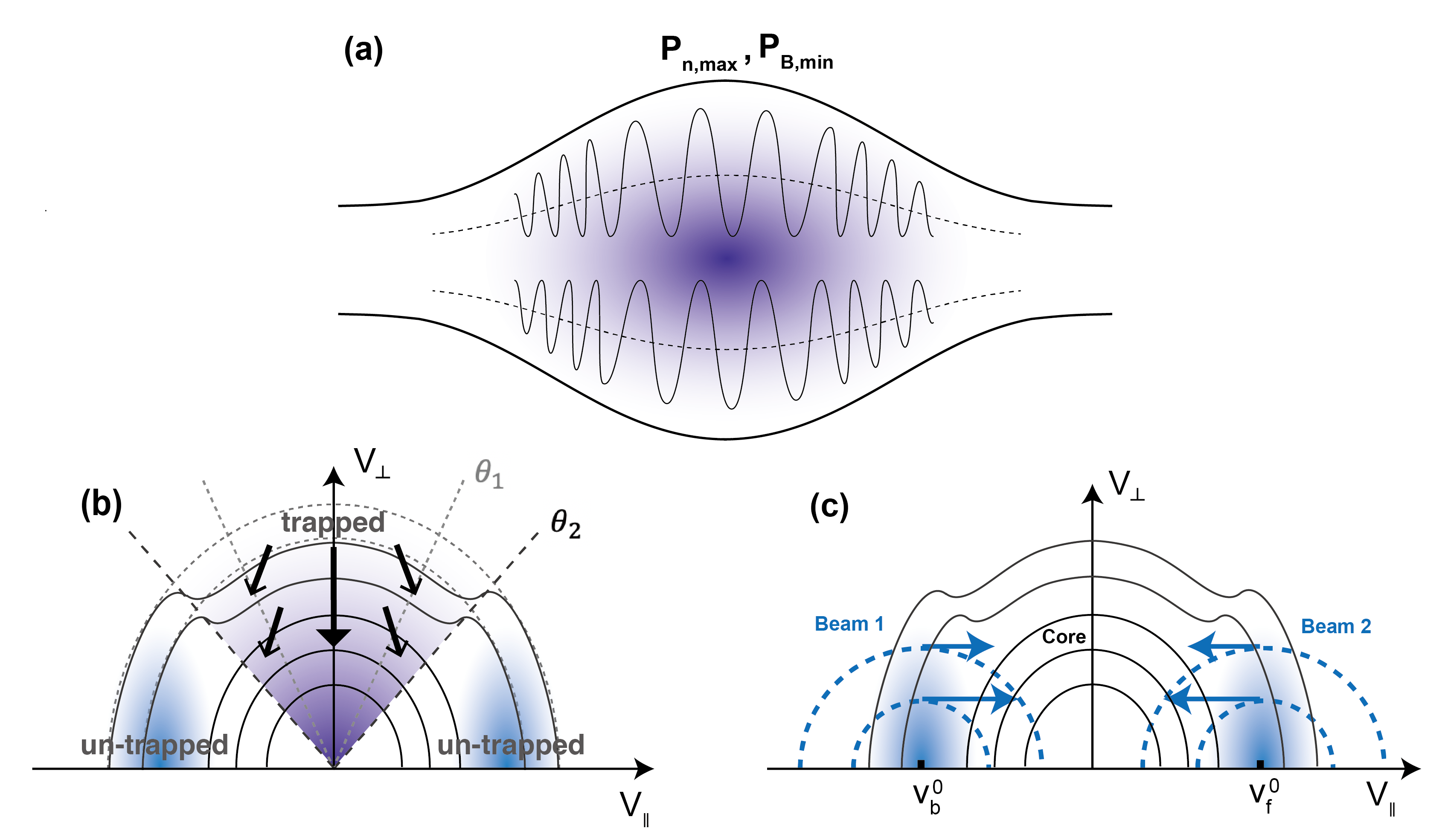}
\caption{Illustration of our conceptual model for a converging magnetic hole. (a) A schematic of the trapped electrons (shaded) in a magnetic hole. The black spirals centered on a dotted line are the bouncing electron trajectories. (b) A sketch of the formation of beams in velocity space by Fermi acceleration and the betatron effect during the convergence of the magnetic hole. Particle velocity-space trajectories at different pitch angles are shown by the black arrows. (c) A sketch of quasi-linear diffusion paths in a two-beam electron velocity distribution function. {The particle diffusion paths (blue arrows) are locally tangent to semi-circles (shown in blue) about the parallel phase speeds $v^\mathrm{0}_\mathrm{f}=\omega_\mathrm{w}/|k_{\parallel}|$ of the forward-propagating waves and $v^\mathrm{0}_\mathrm{b}=-\omega_\mathrm{w}/|k_{\parallel}|$ of the backward-propagating waves. The diffusion paths always point towards lower phase space density.} For (b) and (c), the electron distribution is shown as black semicircles and two beams are highlighted as blue shaded areas.}
\label{fig0}
\end{figure*}

In this section, we develop a conceptual model to explain the multi-scale evolution of a converging magnetic hole. We summarize it visually in Figure \ref{fig0}. 

The magnetic hole is characterised by a spatially non-uniform magnetic field configuration with a significant number of particles trapped near the magnetic field minimum (illustrated as the purple shade in Figure \ref{fig0}a). Due to the non-uniform magnetic field associated with the magnetic hole, particles are subject to the mirror force when their magnetic moment is conserved. The trapping of particles is described by a critical pitch angle $\theta_\mathrm{c}$  \citep{Kivelson1996}, so that 

\begin{equation*}
\begin{split}
\sin \theta_\mathrm{c} =\sqrt{{B}/ {{B}_\mathrm{max}}},
\end{split}
\label{eq1}
\tag{1}
\end{equation*}
where ${B}$ and ${B}_{\mathrm{max}}$ are the local magnetic field and the maximum magnetic field of the structure. Only particles with a pitch-angle $\theta$ that fulfills $\theta_\mathrm{c} < \theta < (180^\circ-\theta_\mathrm{c})$ are effectively trapped and bounce between their mirror points where their velocity component parallel to the magnetic field reverses. This particle motion with the overlaid gyration motion is illustrated by the black spirals in Figure \ref{fig0}a. Particles with a pitch angle $\theta < \theta_\mathrm{c}$ or $\theta > (180^\circ-\theta_\mathrm{c})$ (i.e., inside the loss cone) stream through the magnetic hole. The trapping of particles marks the formation of a non-propagating spatial structure by retaining pressure balance between the local magnetic pressure $P_\mathrm{B}$ and the local plasma thermal pressure $P_\mathrm{n}$ \citep{Schwartz1996,Soucek2011}. If the magnetic hole is not in pressure balance  \citep{Yao2020}, it either converges or diverges until pressure balance is achieved.

In a converging magnetic hole, the convergence of the mirror points of the trapped particles causes type-1 Fermi acceleration due to the conservation of the second adiabatic invariant. The decreasing magnetic field in a converging magnetic hole also causes betatron cooling of particles  \citep{Southwood1993} due to the conservation of the magnetic moment. The Fermi acceleration increases the velocity of the trapped particles in the directions parallel and anti-parallel to the magnetic field, while the betatron cooling decreases the velocity of particles in the perpendicular direction. 

Assuming gyrotropy, Figure \ref{fig0}b shows the isocontours of the velocity distribution function (VDF) $f_{\mathrm{e}}(v_\parallel,v_\perp)$ of particles in the plasma frame in a magnetic hole, where $v_\parallel$ and $v_\perp$ are the velocity components parallel and perpendicular to the magnetic field. The dashed black and grey lines represent different critical pitch angles according to Eq. \ref{eq1} for different local magnetic fields, referred to as $\theta_\mathrm{c1}$ or $180^\circ-\theta_\mathrm{c1}$ and $\theta_\mathrm{c2}$ or $180^\circ-\theta_\mathrm{c2}$. Trapped particles are illustrated with a purple shade in Figure \ref{fig0}b. The black arrows in Figure \ref{fig0}b show the velocity-space trajectories of particles with different pitch angles due to Fermi acceleration and betatron cooling as the magnetic hole converges. 

As a consequence of the Fermi and betatron effects, a pair of counter-streaming electron beams (blue shaded in Figure \ref{fig0}b-c) form in velocity space. Beams represent a non-equilibrium plasma state that can drive kinetic micro-instabilities  \citep{Gary1993}. In particular, counter-streaming electron beams can drive unstable electromagnetic whistler waves with right-handed polarisation and frequencies below the local electron cyclotron frequency. 

Whistler waves are frequently observed in magnetic holes, but the excitation and origin of these waves are a matter of ongoing research  \citep{Zhang1998,Ren2019,Yao2019,Huang2019,Kitamura2020}. Recently, so-called pancake, donut-shaped, or butterfly pitch-angle distributions with beams of electrons have been proposed as sources for the wave excitation possibly via cyclotron resonances  \citep{Zhima2015,Ahmadi2018,Breuillard2018,Behar2020,Huang2020,Zhang2021}. However, there is still a lack of consistent evidence underpining the nature of the resonant waves and the role of the quasi-linear evolution of these instabilities for the evolution of magnetic holes.

In quasi-linear theory, field-aligned electron beams evolve under the action of whistler-wave instabilities via either Landau or cyclotron resonant wave-particle interactions  \citep{Kennel1966,Rowlands1966,Shapiro1962,Verscharen2019a,Jeong2020}. The particles that participate in resonant wave-particle interactions have a velocity component $v_{\parallel}$ parallel to the local magnetic field that fulfills the resonance condition

\begin{equation*}
\begin{split}
v_{\mathrm{\parallel}}=v^\mathrm{n}_{\mathrm{res}}=\frac{\omega_\mathrm{w}+{n}\Omega_{\mathrm{e}}}{k_{\parallel}},
\end{split}
\label{eq2}
\tag{2}
\end{equation*}
where $v^\mathrm{n}_\mathrm{res}$ is the n-th resonance speed, $\omega_\mathrm{w}$ is the real part of the whistler wave frequency, $k_{\parallel}$ is the parallel wavenumber, $n$ is an integer, $\Omega_\mathrm{e}=eB/m_\mathrm{e}$ is the electron gyrofrequency, $e$ is the electron charge, $B$ is the magnetic field, and $m_\mathrm{e}$ is the electron mass. The Landau resonance condition corresponds to $n=0$ in Eq. \ref{eq2}. In that case, only electrons with $v_{\parallel}=v^\mathrm{0}_\mathrm{res}=\omega_\mathrm{w}/k_{\parallel}$ resonate and thus secularly exchange energy with the waves. Meanwhile, resonant electrons undergo diffusion in velocity space along specific trajectories. 

Using a similar format as Figure \ref{fig0}b, we illustrate the Landau resonant interaction between counter-streaming electron beams and unstable whistler waves in Figure \ref{fig0}c. In these interactions, electrons transfer energy to the waves and thus drive them unstable only if they lose energy when undergoing quasi-linear diffusion. Since velocity-space diffusion always occurs from larger values of $f_{\mathrm{e}}$ to smaller values of $f_{\mathrm{e}}$, {this condition requires that $\partial f_{\mathrm{e}}/\partial v_{\parallel}>0$ at $v^\mathrm{0}_\mathrm{f}$ (i.e., forward-propagating waves), and $\partial f_{\mathrm{e}}/\partial v_{\parallel}<0$ at $v^\mathrm{0}_\mathrm{b}$ (i.e., backward-propagating waves).} Unstable slightly-oblique whistler waves with $k_\parallel > 0$ and $k_\parallel < 0$ are excited and the parallel/anti-parallel component of wave electric field is responsible for the particle diffusion. 

In Figure \ref{fig0}c, blue dashed semi-circles illustrate the diffusion paths of resonant particles that undergo wave-particle interactions. The direction of the diffusion is always tangent to semi-circles around the associated $\omega_{\mathrm{w}}/k_{\parallel}$ since quasi-linear diffusion is energy-conserving in the reference frame that moves with the parallel phase speed of the resonant waves  \citep{Verscharen2019b}. We use blue arrows in Figure \ref{fig0}c to illustrate the diffusion trajectories of Landau-resonant electrons in our case. 

For consistency, our multi-scale model requires that the quasi-linear diffusion rate $\nu_{\mathrm{d}}$ is less than the growth rate of the unstable whistler waves $\gamma_{\mathrm{w}}$ which itself must be much less than the wave frequency $\omega_\mathrm{w}$ of the unstable waves. Since the driving of the quasi-linear diffusion depends critically on the trapping effect, we require that $\nu_{\mathrm{d}}$ is less than the trapping frequency $1/\tau_{\mathrm{t}}$ of the bouncing electrons. The slowest process in our model is the ion-scale growth of the magnetic hole, estimated as the linear mirror-mode growth rate $\gamma_{\mathrm{m}}$ which we thus set as the smallest characteristic frequency in our scenario. This time-scale ordering of these processes is given by 

\begin{equation*}
\begin{split}
\gamma_{\mathrm{m}} \ll \nu_{\mathrm{d}} \lesssim \frac{1}{\tau_\mathrm{t}} \ll \gamma_{\mathrm{w}} \ll \omega_{\mathrm{w}} \lesssim \Omega_{\mathrm{e}}.
\end{split}
\label{eq3}
\tag{3}
\end{equation*}

\section{Data set}\label{sec2}
Only recently, direct in-situ measurements of the details of the electron behaviour in magnetic holes on a short time-scale has become possible due to the unprecedented high time-resolution electron velocity distribution data from NASA's Magnetospheric Multi-Scale (MMS) mission. We use data from the MMS mission when the spacecraft were in the Earth's magnetosheath on 2017 January 25 from 00:25:40 UT to 00:26:15 UT. The magnetic field data are provided by the fluxgate magnetometer (FGM)  \citep{Russell2016}. The high time-resolution electromagnetic field data are provided by the search-coil magnetometer (SCM) and the electric double probes (EDP)  \citep{Torbert2016}. The particle velocity distribution data are retrieved by the fast plasma investigation (FPI)  \citep{Pollock2016}. We focus on the high time-resolution dynamics of wave-particle interactions and diffusion of the trapped electrons in the magnetic hole. All data used in this paper are high time-resolution burst mode data.

\section{Results}\label{sec3}
\subsection{Particle trapping and diffusion in the magnetic hole}\label{subsec31}
\begin{figure*}[!htbp]
\centering
\includegraphics[width=1\textwidth]{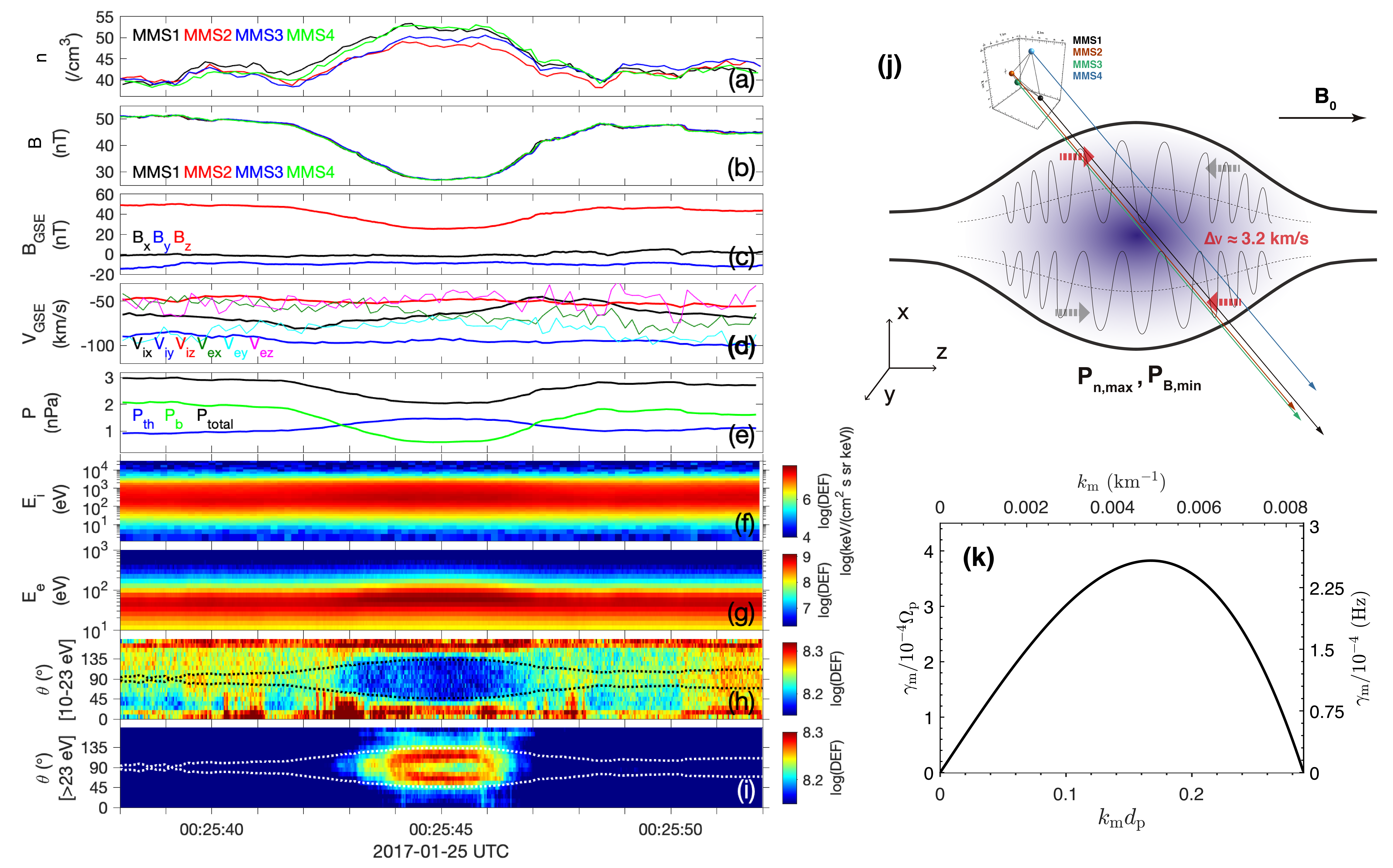}
\caption{MMS observations of a magnetic hole on 2017 January 25 from 00:25:30 UT to 00:26:00 UT. (a) Ion number densities. (b) Magnetic field strengths. (c) Magnetic field components measured by MMS1 in GSE coordinates. (d) Velocity components of ion and electrons measured by MMS1 in GSE coordinates. (e) Ion thermal pressure, magnetic pressure, and the sum of thermal and magnetic pressures. (f) and (g) Ion and electron differential energy flux spectrograms. (h) and (i) Electron pitch-angle spectrograms of $10-23$ eV and $>23$ eV electrons. The black/white dotted lines in (h) and (i) indicate the critical pitch angles according to Eq. \ref{eq1}. (j) A sketch of the magnetic hole crossing by the MMS tetrahedron shown in GSE coordinates. (k) The growth rate of the mirror-mode instability calculated by NHDS using average parameters of the MMS1 measurements.}
\label{fig1}
\end{figure*}

Figures \ref{fig1}a-j show a magnetic hole crossing observed by MMS. Figures \ref{fig1}a and \ref{fig1}b show the ion density and magnetic field strength observed by the four MMS which were in a tetrahedral formation with a quality factor of 0.891 at the time of measurement. The four spacecraft observe almost identical profiles in the measured quantities due to the small spacecraft separation ($\approx$ 10 km) compared to the size of the magnetic hole. We find a clear anti-correlation between the magnetic field strength and the plasma density, which is a characteristic property of magnetic holes. The plasma density assumes its maximum when the magnetic field strength assumes its minimum at the center of the magnetic hole. Figure \ref{fig1}c shows the magnetic field components observed by MMS1 in the Geocentric Solar Ecliptic coordinate system (GSE). The magnetic-field variations are dominated by the field's z-component. Figures \ref{fig1}f and \ref{fig1}g show the differential energy flux of ions and electrons measured by FPI onboard MMS1. The magnetosheath plasma is anisotropic in the magnetic hole and is composed of intense hot ions (few hundreds of eV) and cold electrons (few tens of eV). The electron energy exhibits an local enhancement inside the magnetic hole. 

We calculate the linear growth rate of the mirror-mode instability using the New Hampshire Dispersion Relation Solver (NHDS)  \citep{Verscharen2018b} based on the observed plasma parameters. The average plasma parameters from 00:25:30 UT to 00:26:00 UT (i.e., about three wavelengths of the mirror mode) are: the magnetic field strength ${B}=44.22$ nT,  ion number density $ {n}=43.89 $ cm$^{-3}$, proton perpendicular thermal speed $v_{\mathrm{th,p}\perp}=1.87\times10^2$ km/s, proton parallel thermal speed $v_{\mathrm{th,p}\parallel}=1.44\times10^2$ km/s, electron perpendicular thermal speed $v_{\mathrm{th,e}\perp}=3.03\times10^3$ km/s, electron parallel thermal speed $v_{\mathrm{th,e}\parallel}=2.89\times10^3$ km/s, and plasma bulk speed $v_{\mathrm{sw}}=123.89$ km/s. As shown in Figure \ref{fig1}k, NHDS predicts that the plasma is unstable to the mirror-mode instability with a maximum growth rate of $\gamma_{\mathrm{m}} \approx 0.0003$ Hz at a wave vector of about $k_{\mathrm{m}}=0.005$ km$^{-1}$, corresponding to a wavelength of about $\lambda_{\mathrm{m}}=2\pi/k_{\mathrm{m}}=1257$ km. The angle between the wave number at maximum growth and the magnetic field is $81^\circ$. Since the real frequency of mirror modes is zero (i.e., not propagating), this structure is convected by the plasma bulk flow. The NHDS results for all four MMS spacecraft measurements are almost identical because their separations are much smaller than the structure.

Figure \ref{fig1}j shows a sketch of the MMS trajectory during the magnetic hole crossing. The magnetic hole is elongated along the direction parallel to the magnetic field. The sampling direction relative to the magnetic-field direction of MMS largely depends on the angle between the plasma bulk flow and the background magnetic field ($\theta_\mathrm{bv}$), in this case $\theta_\mathrm{bv}=51^{\circ}$. Thus we approximate the convection time of a half-wavelength of the mirror-mode structure as 

\begin{equation*}
\begin{split}
\tau_\mathrm{c}=\frac{\lambda_{\mathrm{m}} \sin81^\circ}{2 v_{\mathrm{sw}}\sin\theta_\mathrm{bv}} \approx 6.4 s. 
\end{split}
\label{eq31}
\tag{4}
\end{equation*}
This value is in agreement with the duration of the magnetic hole in the MMS observation from 00:25:42 UT to 00:25:48 UT.

Figures \ref{fig1}h and \ref{fig1}i show the pitch-angle distribution functions of electrons with energies 10-23 eV and with energy greater than 23 eV. The black (white) dotted lines in Figures \ref{fig1}h (\ref{fig1}i) represent the critical pitch angle $\theta_\mathrm{c}$ and $180^\circ -\theta_\mathrm{c}$ of particle trapping from Eq. \ref{eq1}. Most of the electrons in these energy ranges are trapped in the magnetic hole (i.e., between both dotted lines). We approximate the typical electron trapping time as 
\begin{equation*}
\begin{split}
\tau_\mathrm{t}=\lambda_{\mathrm{m}}/v_\mathrm{th,\parallel e} \approx 0.42 s, 
\end{split}
\label{eq32}
\tag{5}
\end{equation*}
where $v_\mathrm{th,\parallel e}$ is the local thermal speed parallel to the magnetic field of the electrons at the centre of the magnetic hole. The time scale of the electron bounce motion is much smaller than the time scale of the plasma convection ($\tau_\mathrm{t} \ll \tau_\mathrm{c}$). The depletion of electrons near $90^\circ$ in Figures \ref{fig1}h and \ref{fig1}i is direct evidence for betatron cooling, which are referred to as donut-shaped pitch-angle distributions by recent insitu observations  \citep{Breuillard2018,Ahmadi2018,Soucek2011,Li2021}. As the magnetic hole converges and the magnetic field decreases, betatron cooling reduces the perpendicular velocity of the trapped electrons. In our case, we find a strong depletion of 10-23 eV electrons inside the critical trapping angle in Figure \ref{fig1}h. 

Using the multi-spacecraft timing technique  \citep{Russell1983}, we find that the magnetic hole converges at its inbound boundary with a velocity of about [-0.25, 0.5, -3.25] km/s in GSE coordinates. Due to the pressure gradient, this converging motion almost along the the magnetic field direction directly causes the Fermi acceleration of electrons. This is in agreement with the increase of electron energy in the energy spectrogram shown in Figure \ref{fig1}g. The Fermi acceleration also leads to ``X-shaped'' leakage of $>23$ eV electrons to the loss cone as shown in Figure \ref{fig1}i. All of these signatures are consistent with our predicted large-scale dynamics for a converging magnetic hole illustrated in panels (a) and (b) of Figure \ref{fig0}.

\subsection{Resonant instability of counter-streaming electron beams}\label{subsec32}

\begin{figure*}[!htbp]
\centering
\includegraphics[width=0.9\textwidth]{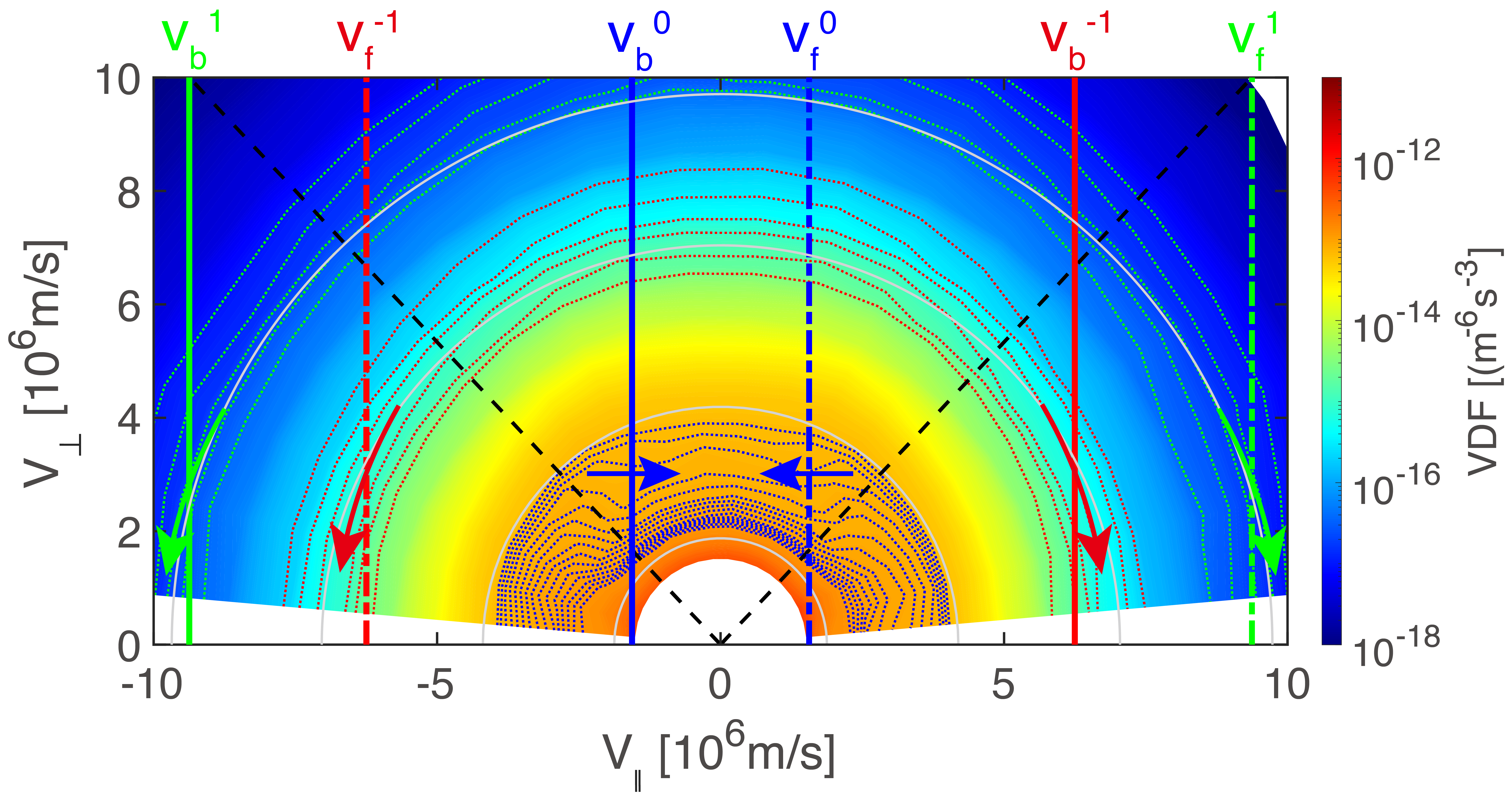}
\caption{Averaged contour plot of the electron VDF measured by MMS1 FPI in the velocity plane ($v_{\mathrm{\parallel}}, v_{\mathrm{\perp}}$) from 2017 January 25 from 00:25:44.38 UT to 00:25:44.80 UT. The blue solid and dashed lines denote the Landau resonance speeds with whistler waves calculated based on the local parameters. {The red and green vertical lines denote the cyclotron resonance speeds with $n=-1$ and $n=1$ in Eq.\ref{eq5}. Dashed and solid line styles represent the resonance speed for the forward-propagating and backward-propagating waves.} The black dashed lines represent the critical pitch angle $\theta_c$ according to Eq.\ref{eq1}. The colored arrows represent the quasi-linear diffusion trajectories of resonant interactions. {The grey semicircles represent constant kinetic energy (i.e., $v_{\perp}^2+v_{\parallel}^2 = \text{constant}$). The colored contour lines are highlited to show the local gradients of the VDF.} VDF values smaller or greater than the colorbar range are not shown here.}
\label{fig2}
\end{figure*}

Figure \ref{fig2} shows the contours of the electron VDF measured by FPI onboard MMS1 near the center of the magnetic hole on 2017 January 25 between 00:25:44.38 UT and 00:26:44.80 UT. The VDFs from all four MMS spacecraft are nearly identical. Errors from photo-electrons and secondary electrons are corrected in the VDFs (see Appendix C). We plot the contours of $f_{\mathrm{e}}(v_{\parallel},v_{\perp})$ in a layout similar to Figure \ref{fig0}b. The black dashed lines denote the critical pitch angles $\theta_\mathrm{c}$ and $180^\circ-\theta_\mathrm{c}$ according to Eq.\ref{eq1}. {The blue vertical lines show the Landau resonance speeds of the forward-propagating (dashed) whistler waves with phase speed $v^\mathrm{0}_\mathrm{f}$ and the backward-propagating (solid) whistler waves with phase speed $v^\mathrm{0}_\mathrm{b}$ for which the integer $n$ in Eq.\ref{eq2} is equal to 0. The red vertical line represents the cyclotron resonance speeds for $n=-1$ in Eq.\ref{eq2} for forward-propagating (dashed) and backward-propagating (solid) whistler waves. The green lines mark the same for $n=+1$ in Eq.\ref{eq2} for forward-propagating (dashed) and backward-propagating (solid) whistler waves.}

The electron VDF is non-Maxwellian and has two significant enhancements near ${v_\parallel}\approx \pm 2\times10^{3}$ km/s, corresponding to a pair of counter-streaming electron beams with energies between 10-23 eV. This pair of electron beams is partially located just outside the critical trapping pitch angle (i.e., inside the loss cone), consistent with our prediction shown in Fugure \ref{fig0}b. 

We calculate the Landau resonance speed of forward-propagating (backward-propagating) whistler waves $v^\mathrm{0}_\mathrm{f}={1.8\times 10^3}$ ($v^\mathrm{0}_\mathrm{b}={-1.8\times 10^3}$) km/s {by inserting $n=0$ to}
\begin{equation*}
\begin{split}
{v^\mathrm{n}_\mathrm{b/f}}=\left(\frac{B^2}{4\mu_0 m_\mathrm{e} n_\mathrm{e}} \frac{\Omega_\mathrm{e}}{\omega_\mathrm{w} \cos^2\theta_\mathrm{k}}\left( \cos \theta_\mathrm{k}-\frac{\omega_\mathrm{w}}{\Omega_\mathrm{e}}\right)\left( n+\frac{\omega_\mathrm{w}}{\Omega_\mathrm{e}}\right)\right)^{\frac{1}{2}},
\end{split}
\label{eq5}
\tag{6}
\end{equation*}
which is a quasi-linear approximation based on the cold-plasma dispersion relation  \citep{LengyelFrey1994}, where $\mu_0$ is the vacuum permeability. We use a wave angle $\theta_\mathrm{k}=10^\circ$ and a ratio between the whistler waves and electron gyro-frequency of $\omega_\mathrm{w}/\Omega_\mathrm{e}= 0.3$ since this is close to the frequency of whistler waves in our observation (see Section \ref{subsec33} for more details). 

We find that $\partial f_{\mathrm{e}}/\partial v_{\parallel}>0$ at $v^\mathrm{0}_\mathrm{f}=1.8\times 10^3$ km/s and $\partial f_{\mathrm{e}}/\partial v_{\parallel}<0$ at $v^\mathrm{0}_\mathrm{b}=-1.8\times 10^3$ km/s, meaning that Landau resonance with the electric-field component $E_{\parallel}$ parallel to the background magnetic field leads to an instability of the resonant waves. If the whistler waves are not exactly parallel, they have a sufficient amplitude in $E_{\parallel}$ and can thus participate in Landau-resonant wave-particle interactions. This result strongly suggests that the counter-streaming electron beams are the local driver for the whistler wave generation in the magnetic hole. During the whistler-wave generation, the unstable electron VDF diffuses as shown in panel (c) of Figure \ref{fig0} of our conceptual model. Like in our Figure \ref{fig0}c, we use the blue arrows to show the quasi-linear diffusion trajectories of the resonant wave-particle interaction in our Figure \ref{fig2}.

{Depending on the local gradient of the electron VDF at the cyclotron resonance speeds, cyclotron resonant interactions also contribute to the growth/damping of the whistler waves. We use the red and green arrows in Figure \ref{fig2} to show possible diffusion paths of cyclotron resonant electrons in velocity space. In our particular example, the cyclotron-resonant electrons with $n=-1$ decrease in thier kinetic energy if the diffusive particle flux in velocity space is directed along the green arrows shown in Figure \ref{fig2}. This cyclotron-resonant interaction contributes to the growth of the resonant whistler waves. In the case of forward-propagating whistler waves with $\omega_\mathrm{w}/k_{\parallel}>0$, the only available resonance for electron-wave interactions is the cyclotron resonance with $n=-1$. In the case of backward-propagating whistler waves with $\omega_\mathrm{w}/k_{\parallel}<0$, it is only the resonance with $n=1$. However, in the case of the resonance with $n=1$, the direction of the diffusive particle flux is towards greater ($v_{\perp}^2+v_{\parallel}^2$) due to the gradients of the electron VDF at the resonance speed, which thus corresponds to a contribution to the damping of the resonant whistler waves. The overall instability of the whistler waves is the result of contributions from resonances with all accessible $n$.}

\subsection{Properties of the unstable whistler waves: theory and observations}\label{subsec33}

We mathematically evaluate the consequence of the large-scale convergence: the stability of the observed electron VDFs in the magnetic hole, with the Arbitrary Linear Plasma Solver (ALPS) code \citep{Verscharen2018a}. We model the electron VDF by using a four-component VDF model, which has two drifting electron beams, one Maxwellian core, and one modified Moyal distribution to model the observed flat-top distribution  (see Apendix B for more details). Different from traditional Maxwellian plasma dispersion solvers, ALPS allows us to include the non-Maxwellian Moyal component. We implement a fitting technique for the electron VDF data around the centre of the whistler-wave emission on 2017 January 25 from 00:25:44.38 UT to 00:26:44.80 UT. The fit parameters for the electron VDF shown in Figure \ref{fig2} are listed in Table \ref{tab1}. {The electron plasma frequency is about 12.1$\Omega_\mathrm{e}$ (i.e., $5.87\times10^{4}$ Hz).}

\begin{figure*}[!htbp]
\centering
\includegraphics[width=0.99\textwidth]{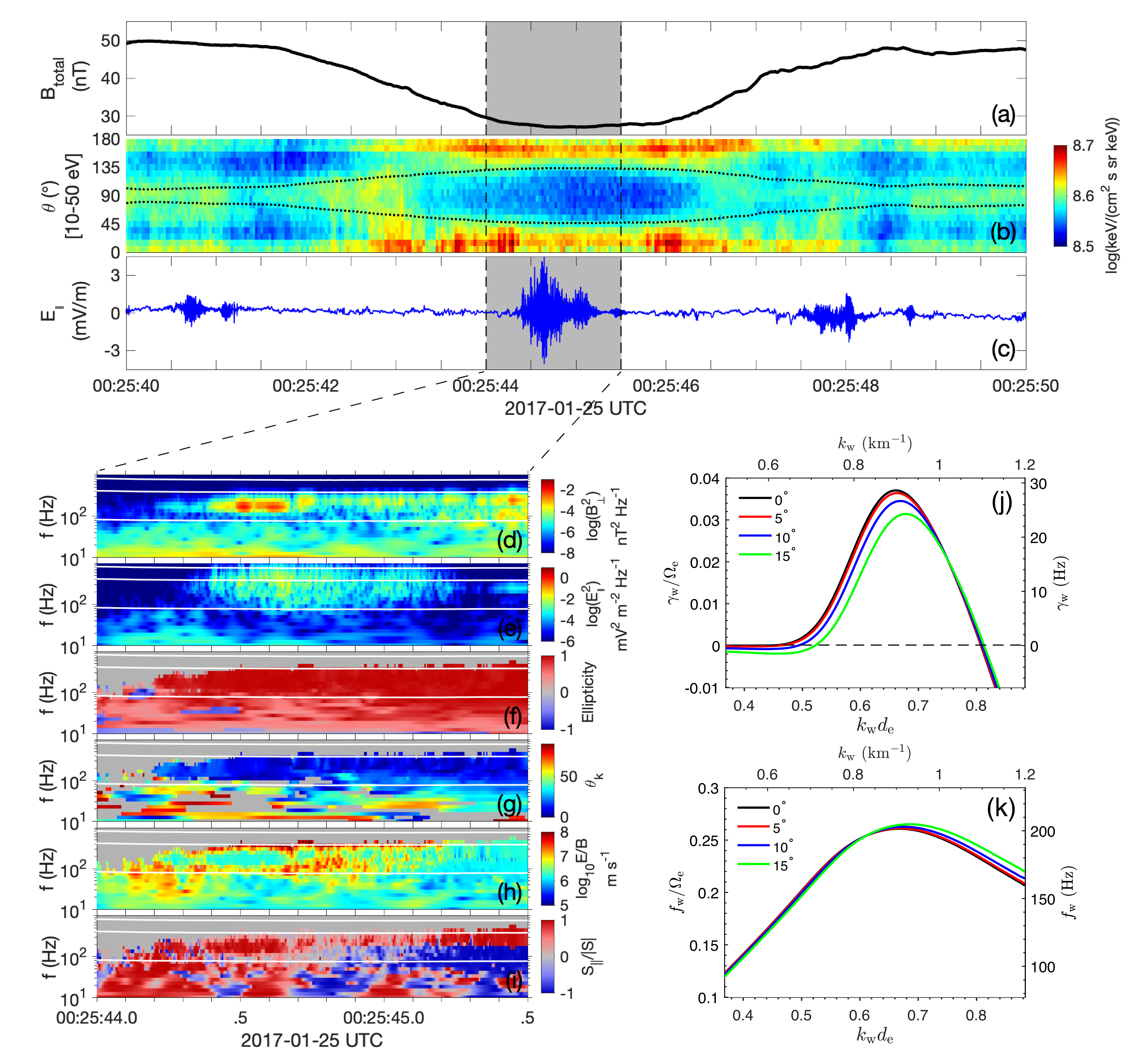}
\caption{Polarization analysis and numerical prediction of the unstable whistler waves from MMS1. (a) the magnetic field strength. (b) pitch-angle distribution of electrons from FPI with energy within 10-50 eV. (c) the parallel electric field from EDP. (d) the power spectral density of the magnetic field fluctuations perpendicular to the background from SCM. (e) the power spectral density of the electric field fluctuations parallel to the background from EDP. (f) the ellipticity. (g) the wave angle. (h) the phase speed. (i) the Poynting flux. The three white solid lines in (d)-(i) represent frequencies corresponding to $0.1 \Omega_{\mathrm{e}}$, $0.5 \Omega_{\mathrm{e}}$ and $\Omega_{\mathrm{e}}$. (j) and (k) our model predictions of the growth rate and the frequency of unstable whistler-mode with different wave angles calculated by ALPS.}
\label{fig3}
\end{figure*}

The growth rate and real frequency of the unstable whistler-mode predicted by our ALPS calculations are shown in Figures \ref{fig3}j and \ref{fig3}k. With a wave angle of $\theta_\mathrm{k}=10^\circ$, the whistler wave has a growth rate of $\gamma_\mathrm{w}=27$ Hz and a real frequency of $f_\mathrm{w} \approx 203$ Hz at a wave vector of $k_\mathrm{w}=0.81$ km$^{-1}$. We find the corresponding phase speed of $v_\mathrm{w}=2\pi f_\mathrm{w}/k_\mathrm{w} \approx 1.56\times10^{3}$ km/s.

We show the observed whistler waves with the help of high-cadence data of electric and magnetic fields measured by the SCM and EDP instruments onboard MMS1. We use the electric field fluctuations in a field-aligned coordinate system based on the average background magnetic field. Figure \ref{fig3}a-c shows that the electric field fluctuations parallel to the magnetic field $E_{\parallel}$ have a significant enhancement near the magnetic field minimum at the center of the magnetic hole. The parallel electric field enhancement in the grey-shaded region coincides with the VDF shown in Figure \ref{fig2}.

Consistently with our multi-scale ordering and the required properties of resonant whistler waves, we find that slightly oblique whistler waves exist at the magnetic hole center. Figure \ref{fig3}d-i show the polarization properties of the detected electro-magnetic field fluctuations. We use the singular value decomposition (SVD) method to calculate the power spectral densities of the perpendicular magnetic field fluctuations, the parallel electric field fluctuations, and the polarization properties of the wave fields, such as the ellipticity, the wave angle $\theta_\mathrm{k}$, the phase speed and the Poynting flux \citep{Santolik2003}. 

As shown in Figures \ref{fig3}d-e, we find significant enhancements of the perpendicular magnetic field and the parallel electric field fluctuations at about 202 Hz (i.e., about 0.26 $\Omega_\mathrm{e}$) near the center of the magnetic hole on 2017 January 25 between 00:25:44.48 UT and 00:26:44.54 UT. Within this interval, the ellipticity of the wave and the degree of polarization is close to unity, suggesting a slightly oblique right-hand polarized wave. The phase speed averaged at the frequency of 202 Hz during this interval is about $1.58\times10^{3}$ km/s. The wave propagates with an average angle of $\theta_{\mathrm{k}}\approx 9^\circ$ with respect to the direction parallel or anti-parallel to the magnetic field. The whistler waves propagate away from the source and predominantly along the field direction. Since the counter-streaming electron beams are not exactly symmetric, we observe an unbalanced Poynting flux in the waves. These observations are in good agreement with the predictions of our ALPS calculations. 

We estimate the diffusion rate $\nu_\mathrm{d} $ based on the quasi-linear approximation 

\begin{equation*}
\begin{split}
\nu_\mathrm{d} \approx \frac{c^2 \Omega_\mathrm{e}^2}{2 B_0^2} \frac{v_\mathrm{sw} k_\mathrm{w}^3}{\omega_\mathrm{w}^3} \hat E_\parallel^2 (\omega_\mathrm{w}) \cos\theta_\mathrm{bv} = 0.25 \mathrm{Hz},
\end{split}
\label{eq6}
\tag{7}
\end{equation*}
where $c$ is the speed of light and $\hat E_\parallel^2$ is the power spectral density of the component of electric field fluctuations parallel to the magnetic field at a frequency of 202 Hz (see Apendix A for more details). As required in Eq.\ref{eq2}, the diffusion rate lies between the fast whistler wave growth ($\gamma_\mathrm{w} \approx$ 27 Hz) and the slow mirror instability growth ($\gamma_\mathrm{m} \approx$ 0.0003 Hz). The wavelength of the whistler wave is about $\lambda_\mathrm{w}=2\pi/k_\mathrm{w} \approx 7$ km, which is equivalent to about 10$\rho_\mathrm{e}$. We note this characteristic length is much smaller than the length scale of the magnetic hole. {This scale ordering is consistent with our scenario that the whistler waves are locally generated and diffuse electrons via Landau resonant interactions at the local gradients of the VDF shown in Figure \ref{fig0} and \ref{fig2}.}

\section{Discussion and Conclusions}\label{sec4}
We show clear in-situ evidence for Landau resonant wave-particle interaction between electrons and slightly oblique whistler waves in a converging magnetic hole in the Earth's magnetosheath. We propose a conceptual model and a consistent ordering for the multi-scale particle dynamics in such a structure. In this model, a converging magnetic hole generates whistler waves through the interplay between Fermi acceleration, betatron cooling, and {resonant wave-particle interactions}. 

We test this idea with MMS observations of a magnetic hole that converges with a velocity of [-0.25, 0.5, -3.25] km/s mostly in the direction of the background magnetic field. As proposed in our model, a pair of counter-streaming electron beams is produced. 

{We observe and explain the local generation of whistler waves at the magnetic-hole center by a combination of Landau-resonant and cyclotron-resonant wave-particle interactions. The Landau-resonant interaction is of particular interest for our understanding of the evolution of the magnetic hole. It fills the velocity space between the electron beams and thus effectively smoothes out the non-thermal features created by Fermi acceleration and betatron cooling.}

The studied magnetic hole fulfils the ordering of scales from Eq. \ref{eq3}: (1) The maximum growth rate of the mirror-mode instability is $\gamma_\mathrm{m}\approx $0.0003 Hz at a wavelength of $\approx$ 1257 km. (2) The estimated quasi-linear diffusion rate for the Landau resonant wave-particle interaction of electrons is 0.25 Hz. (3) The typical trapping frequency of electrons is 2.38 Hz. (4) The growth rate of the unstable whistler waves is 30 Hz. (5) The frequency of of the unstable whistler waves is about 202 Hz. In this ordering, the Landau resonant wave-particle interaction secularly transfers the kinetic energy of electrons to the unstable whistler waves on a time scale greater than the typical time of the electron bounce motion in the magnetic hole structure. 

Our study develops and confirms a consistent understanding of the evolution of converging magnetic holes. Magnetic holes are an important type of coherent structure in space plasmas that evolve through a multi-scale process that couples the kinetic dynamics of particle diffusion and energy transfer at electron scales. As a remarkable signature of these converging magnetic holes, we find that whistler waves are an important feature of local energy emission from the kinetic energy of resonant electrons. We note that the saturation and non-linear development of the mirror-mode instability play a non-trivial role in the particle diffusion in magnetic holes \citep[e.g.,][]{Kivelson1996,Kuznetsov2007}. As a type of quasi-steady plasma structure, magnetic holes experience different stages of their evolution, in some of which electron-scale waves are present or absent \citep[e.g.,][]{Ahmadi2018,Huang2019}. {Figure 4e shows the existence of electrostatic fluctuations above the frequency of our whistler waves. Multiple forms of kinetic waves such as electrostatic solitary waves and electron cyclotron waves have been observed in magnetic holes \citep[e.g.,][]{Yao2019}. However, these additional wave-particle mechanisms associated with magnetic holes remain beyond the scope of our current analysis.} 

Building on our model and observations, it would be worthwhile to undertake a full numerical evaluation of the adiabatic effects and the resonant evolution in a quasi-linear framework. This would further improve our understanding of the multi-scale particle dynamics in coherent structures.

\acknowledgments
We thank the MMS team for data. All of the data used in this paper are accessible at the MMS Science Data Center (\url{https://lasp.colorado.edu/mms/sdc/public}). The code for data and polarization analysis is publically available via \url{https://github.com/irfu}. More description on the Arbitrary Linear Plasma Solver (ALPS) and the New Hampshire Dispersion Relation Solver (NHDS) can be found via \url{http://www.alps.space} and \url{https://github.com/danielver02/NHDS}. W.J. thanks Dr. Wenya Li, Dr. Bingbing Tang, Dr. Seong-Yeop Jeong, Dr. Yong Ren, Dr. Chen Zeng, Chuanhui Gao and Xingyu Zhu for helpful discussions and suggestions. This work is supported by NNSFC grants (42022032, 41874203, 42188101), project of Civil Aerospace ``13th Five Year Plan" Preliminary Research in Space Science (D020301, D030202), Strategic Priority Research Program of CAS (Grant No.XDA17010301), and Key Research Program of Frontier Sciences CAS (Grant No. QYZDJ-SSW-JSC028). W.J. is supported by the CAS Joint Ph.D. training program during his stay at UCL's MSSL. D.V. is supported by the STFC Ernest Rutherford Fellowship ST/P003826/1 and STFC Consolidated Grants ST/S000240/1 and ST/W001004/1. H.L. is supported by International Partnership Program of CAS (Grant No. 183311KYSB20200017) and in part by the Specialized Research Fund for State Key Laboratories of China. K.G.K. is supported by grant DE-SC0020132.

\appendix
\section{Estimate for the quasi-linear diffusion rate}\label{subsec42}

In quasi-linear theory, the Landau resonant wave-particle interaction between electrons and slightly oblique whistler waves leads to a slow (compared to the wave period) time evolution of the VDF according to  \citep{Kennel1966,Marsch2006}
\begin{equation*}
\begin{split}
\frac{\partial f_{\mathrm{e}}}{\partial t} \approx \frac{\mathrm{e}^2}{8 \pi^2 m_\mathrm{e}^2 V}\int \left( \frac{k_\parallel}{\omega_\mathrm{w}}\right)^2 \tilde G \frac{1}{\lvert v_\parallel \rvert} \delta{\left(\omega_\mathrm{w}-k_\parallel v_\parallel \right)}\lvert \psi \rvert^2 \tilde G f_\mathrm{e} d^3 k ,
\end{split}
\tag{A1}
\label{A1}
\end{equation*}
where $V$ is the volume in which the wave amplitude is effective to cause wave-particle interactions,
\begin{equation*}
\begin{split}
\tilde G \approx \frac{k_\parallel v_\perp}{\omega_\mathrm{w}}\frac{\partial}{\partial v_\parallel},
\end{split}
\tag{A2}
\label{A2}
\end{equation*}

\begin{equation*}
\begin{split}
\psi \approx \frac{v_\parallel}{v_\perp} \tilde E_\mathrm{z}, 
\end{split}
\tag{A3}
\label{A3}
\end{equation*}
and $\tilde E_\mathrm{z}$ is the Fourier amplitude of the parallel component of the electric-field fluctuations. Resolving the $\delta$-function in Eq.\ref{A1}, we obtain the simplified quasi-linear diffusion equation:

\begin{equation*}
\begin{split}
\frac{\partial f_{\mathrm{e}}}{\partial t} \approx \frac{{e}^2}{8 \pi^2 m_\mathrm{e}^2 V} \frac{\partial}{\partial v_\parallel}\int \frac{1 }{\lvert v_\parallel\rvert} \bigg \lvert \tilde E_\mathrm{z} \left( k_\parallel = \frac{\omega_\mathrm{w}}{v_\parallel}\right)\bigg \rvert^2 \frac{\partial f_{\mathrm{e}}}{\partial v_\parallel} dk_xdk_y,
\end{split}
\tag{A4}
\label{A4}
\end{equation*}
where $k_x$ and $k_y$ are perpendicular components of the wave vector $k$. Assuming that the electric-field fluctuations have a significant effect on a finite range of $k_{\parallel}$ values with width $\Delta k_{\parallel}$, we find 

\begin{equation*}
\begin{split}
\int \lvert \tilde E_\mathrm{z} \rvert^2 dk_xdk_y = \frac{8 \pi^3 V}{\Delta k_\parallel} \delta E^2, 
\end{split}
\tag{A5}
\label{A5}
\end{equation*}
where $\Delta k_{\parallel}$ is a finite parallel wave number range in which the wave power $\delta E^2$ is distributed. We appoximate $\delta E^2$ of the observed whistler waves by using the power spectral density of the electric-field fluctuations $\hat E_\parallel^2 (\omega)$ in a narrow range of frequency $\Delta \omega_\mathrm{w}$ in the spacecraft frame according to

\begin{equation*}
\begin{split}
\delta E^2=\int \hat E_\parallel^2 (\omega) d\omega \approx \hat E_\parallel^2 (\omega_\mathrm{w}) \Delta \omega_\mathrm{w}.
\end{split}
\tag{A6}
\label{A6}
\end{equation*}

We then use Taylor's hypothesis  \citep{Taylor1938}

\begin{equation*}
\begin{split}
2\pi \Delta \omega \approx v_\mathrm{sw} \cos\theta_\mathrm{bv} \Delta k_\parallel,
\end{split}
\tag{A7}
\label{A7}
\end{equation*}
to express

\begin{equation*}
\begin{split}
\int \lvert \tilde E_\mathrm{z} \rvert^2 dk_xdk_y = 4 \pi ^2 V \cos\theta_\mathrm{bv} v_\mathrm{sw} \hat E_\parallel^2 (\omega_\mathrm{w}).
\end{split}
\tag{A8}
\label{A8}
\end{equation*}

This allows us to estimate the effect of the quasi-linear diffusion as 

\begin{equation*}
\begin{split}
\frac{\partial f_{\mathrm{e}}}{\partial t} \approx  \frac{\partial^2}{\partial v_\parallel^2} \left(\widetilde D f_\mathrm{e} \right) ,
\end{split}
\tag{A9}
\label{A9}
\end{equation*}
where 

\begin{equation*}
\begin{split}
\widetilde D = \frac{\omega_\mathrm{w}^2}{k_\mathrm{w}^2} \nu_\mathrm{d}
\end{split}
\tag{A10}
\label{A10}
\end{equation*}
is the effective diffusion coefficient and

\begin{equation*}
\begin{split}
\nu_\mathrm{d} \approx \frac{c^2 \Omega_\mathrm{e}^2}{2 B_0^2} \frac{v_\mathrm{sw} k_\mathrm{w}^3}{\omega_\mathrm{w}^3} \hat E_\parallel^2 (\omega_\mathrm{w}) \cos\theta_\mathrm{bv}
\end{split}
\tag{A11}
\label{A11}
\end{equation*}
is the estimated diffusion rate. Based on the solutions $\omega_{\mathrm w}$ and $k_{\mathrm w}$ from our linear-theory results and with $\hat E_\parallel^2 \approx 6\times 10^{-4}$ $\mathrm{mV^2m^{-2}Hz^{-1}}$ from our observations at 202\,Hz, we obtain the diffusion rate of $\nu_\mathrm{d} \approx 0.25$ Hz.

\section{VDF model and the Arbitrary Linear Plasma Solver (ALPS)}\label{subsec43}
The measured distribution of the trapped particles in magnetic holes deviates from the equilibrium Maxwellian distribution, especially due to the presence of electron beams and the flat-top part of the distribution, which we model with a bi-Moyal distribution. The bi-Moyal distribution is a two-dimensional extension of the modified Moyal distribution  \citep{Klein2016}. 

To solve the full hot-plasma dispersion relation, we implement a VDF fitting model as input to our ALPS solver. Our VDF model is a combination of three bi-Maxwellian distributions (two beams $f_\mathrm{b}$ and one core $f_\mathrm{c}$) and one bi-Moyal distribution $f_\mathrm{M}$: 

\begin{equation*}
\begin{split}
f_{\mathrm{e}}=f_{\mathrm{c}}+f_{\mathrm{b}}+f_{\mathrm{M}},
\end{split}
\tag{B1}
\label{B1}
\end{equation*}
where 
\begin{equation*}
\begin{split}
f_{\mathrm{b}}(v_{\parallel},v_{\bot})=& \sum_{{\mathrm{j}}=1}^2 \frac{n_{\mathrm{j}}}{\pi^{3/2} v_{{\mathrm{th}}\bot {\mathrm{j}}}^2 v_{{\mathrm{th}}\parallel {\mathrm{j}}}} \exp{\left(-\frac{(v_\bot-u_{\bot {\mathrm{j}}})^2}{v_{{\mathrm{th}}\perp {\mathrm{j}}}^2} -\frac{(v_{\parallel}-u_{\parallel {\mathrm{j}}} )^2}{v_{{\mathrm{th}}\parallel {\mathrm{j}}}^2}\right) };
\end{split}
\tag{B2}
\end{equation*}

\begin{equation*}
\begin{split}
f_{\mathrm{c}}(v_{\parallel},v_{\bot})=\frac{n_3}{\pi^{3/2}v_{{\mathrm{th}}\bot3}^2 v_{{\mathrm{th}}\parallel3}}\exp{\left(-\frac{v_\bot^2}{v_{{\mathrm{th}}\bot3}^2} 
- \frac{(v_\parallel-u_{\parallel3})^2}{v_{{\mathrm{th}}\parallel3}^2}\right)}
\end{split}
\tag{B3}
\end{equation*}
and 

\begin{equation*}
\begin{split}
f_{\mathrm{M}}(v_{\parallel},v_{\bot})=A\,\exp \left(\frac{1}{2} \left[\frac{v_\bot^2}{v_{{\mathrm{th}}\bot \mathrm{M}}^2}+\frac{v_\parallel^2}{v_{{\mathrm{th}}\parallel \mathrm{M}}^2}-\exp \left(\frac{v_\bot^2}{v_{{\mathrm{th}}\bot \mathrm{M}}^2}+\frac{v_\parallel^2}{v_{{\mathrm{th}}\parallel \mathrm{M}}^2}\right)\right]\right),
\end{split}
\tag{B4}
\end{equation*}
where $n_\mathrm{j}$, $v_{{\mathrm{th}}\mathrm{\bot j}}$, $v_{\mathrm{{\mathrm{th}}\parallel j}}$,$v_{\mathrm{{\mathrm{th}}\bot M}}$, $v_{\mathrm{{\mathrm{th}}\parallel M}}$, $u_{\bot \mathrm{j}}$, $u_{\parallel \mathrm{j}}$ and $A$ are fit parameters. By using the Levenberg-Marquardt fitting technique \citep{Press1996}, we obtain model parameters for the total electron VDF by minimizing the residual error. 

\begin{figure*}[!htbp]
\centering
\includegraphics[width=0.9\textwidth]{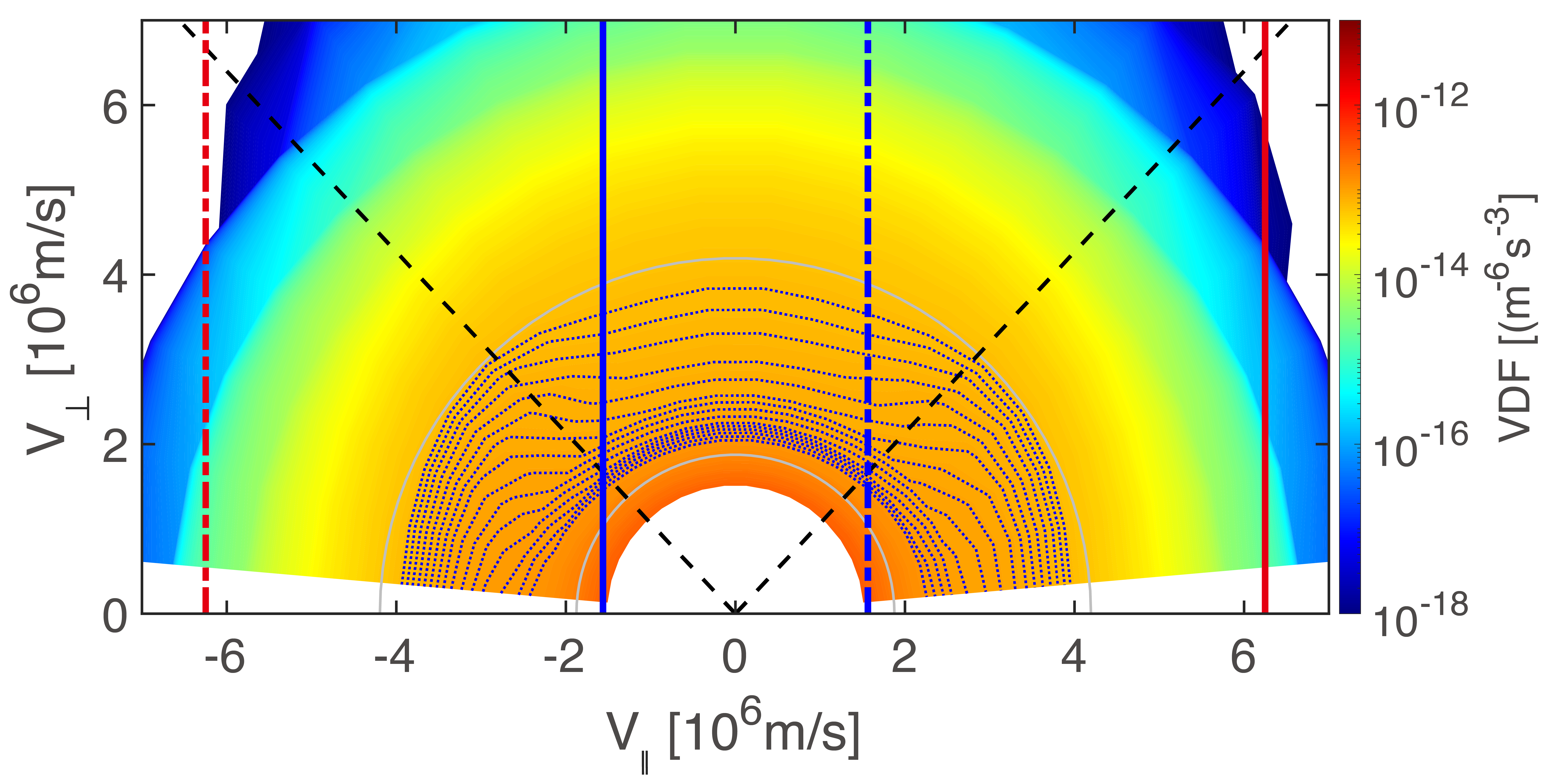}
\caption{Contour plot of our VDF model according to Eq. \ref{B1} based on the averaged electron VDF data from 2017 January 25 from 00:25:44.38 to 00:26:44.80 UT observed by MMS1. The format is the same as Figure \ref{fig2}.}
\label{fig4}
\end{figure*}

We show the obtained parameters for our VDF model in Table \ref{tab1} and the corresponding contour of VDF model in Figure \ref{fig4}. The minimized sum of squared residuals is only 0.04 on a logarithmic scale, indicating a good approximation to the real MMS VDF data. In fact, by comparing to previous models with only bi-Maxwellian components  \citep[e.g.,][]{Huang2020,Zhang2021}, we find that our model is a simple and realistic description.

\begin{table}[h]
\begin{center}
\begin{minipage}{\textwidth}
\caption{The fitting parameters of our VDF data shown in Figure \ref{fig2} on 2017 January 25 from 00:25:44.38 UT to 00:26:44.80 UT.}\label{tab1}
\begin{tabular*}{\textwidth}{@{\extracolsep{\fill}}lccccc@{\extracolsep{\fill}}}
\toprule%

& $n_\mathrm{j}$ (m$^{-3}$)    & $v_{\mathrm{th}\perp \mathrm{j}} $ (m/s) & $v_{\mathrm{th}\parallel \mathrm{j}}$ (m/s) & $v_{\parallel \mathrm{j}} $ (m/s) & $v_{\perp \mathrm{j}}$ (m/s) \\
\midrule
Beam 1 & $1.33\times10^{6}$    & $2.77\times10^{6}$        & $1.17\times10^{6}$            & $2.99\times10^{6}$        & $3.99\times10^{5}$        \\ 
Beam 2 & $2.51\times10^{6}$   & $3.27\times10^{6}$        & $1.05\times10^{6}$            & $-2.86\times10^{6}$       & $-6.35\times10^{5}$       \\ 
Core   & $1.06\times10^{7}$   & $1.11\times10^{6}$        & $1.16\times10^{6}$            & $8.71\times10^{3}$        &               \\ 
\toprule
       & $A$ (m$^{-6}$s$^{-3}$) & $v_{\mathrm{th}\perp \mathrm{M}}$ (m/s)        & $v_{\mathrm{th}\parallel \mathrm{M}}$ (m/s)             &               &               \\ 
\midrule
Moyal  & $1.39\times10^{-13}$ & $4.52\times10^{6}$        & $3.91\times10^{6}$            &               &               \\ 
\botrule
\end{tabular*}
\end{minipage}
\end{center}
\end{table} 

\section{Impact of photo-electrons and secondary electrons}\label{subsec44}
Photo-electrons and secondary electrons are the main sources of error in the measurement of the electron velocity distribution function. Electrons are produced when EUV photons strike a spacecraft surface or an instrument, resulting in both internal and external errors in electron measurements. The finite spacecraft potential $ \phi $ accelerates low-energy electrons and modifies the measurements in the energy range $\lesssim$ $\lvert e\phi \rvert$. MMS has a spacecraft potential controller (ASPOC) that can actively decrease the absolute spacecraft potential and consequently the energy of spacecraft photo-electrons. 

\begin{figure*}[!htbp]
\centering
\includegraphics[width=0.99\textwidth]{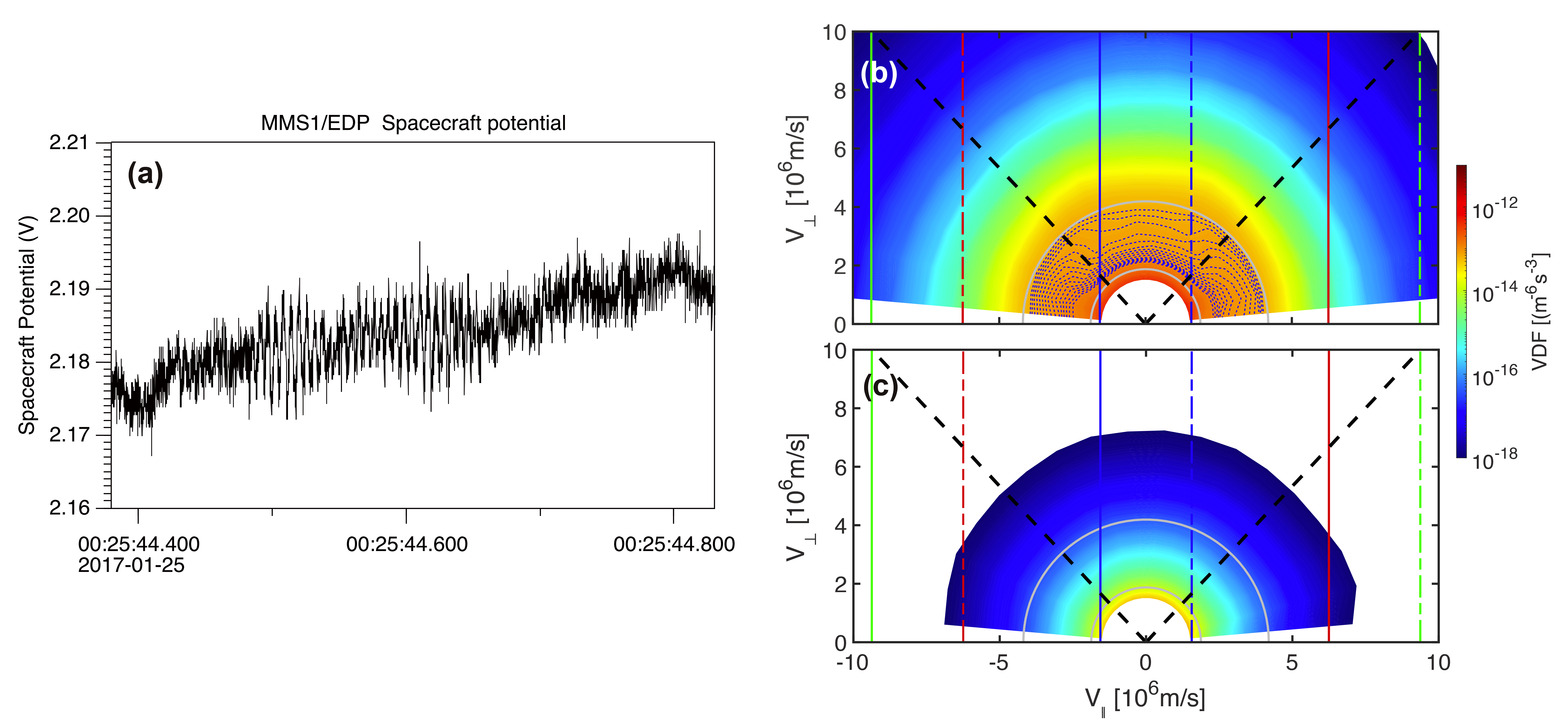}
\caption{Errors of electron velocity distribution function. (a) MMS1 EDP spacecraft potential. (b) the contour plot of the electron VDF without secondary electrons removed. (c) the contour plot of the secondary electron VDF. The VDF time intervals selected in (b) and (c) are the same as in Figure \ref{fig2}.}
\label{fig5}
\end{figure*}

Figure \ref{fig5}a shows the spacecraft potential which is $\approx 2.18$\,V during our measurement interval, suggesting that the error from spacecraft photoelectrons is negligible in our study. 

When the instrument faces the Sun, secondary electrons are produced inside the instrument and independent of spacecraft potential. Figures \ref{fig5}b and \ref{fig5}c show the electron VDF without correction and the VDF of secondary electrons being corrected. We apply this correction throughout our analysis. More details about the correction are given at \url{https://lasp.colorado.edu/galaxy/display/MFDPG/DES+Photoelectrons+-+further+details}.

\end{document}